\documentstyle[prl,aps]{revtex}
\begin{document}
\preprint{LPTENS-96/18}
\draft
\title{Dynamics of particles and manifolds in a quenched
random force field}
\author{Pierre Le Doussal$^{\dag}$,
Leticia F. Cugliandolo$^{\ddag}$,
and Luca Peliti$^{\P}$}
\address{$^{\dag}$ CNRS-Laboratoire de Physique Th\'eorique de l'Ecole\\
Normale Sup\'erieure, 24 rue Lhomond, F-75231 Paris\cite{frad}\\
$^{\ddag}$ Laboratoire de Physique Th\'eorique des Liquides, Jussieu\\
4, Place Jussieu,
75005 Paris France
 \cite{junk}\\
$^{\P }$ 
Groupe de Physico-Chimie Th\'eorique, CNRS URA 1382,
ESPCI, 10, rue Vauquelin, F-75231 Paris Cedex 05, France,\cite{junk2}\\
and\\
Dipartimento di Scienze Fisiche, Unit\`a INFM,
Universit\`a ``Federico II'', Mostra d'Oltremare, Pad.~19,
I-80125 Napoli, Italy.}

\date{\today}
\maketitle

\begin{abstract}
We study the dynamics of a directed manifold
of internal dimension $D$ in a 
$d$-dimensional random force field.
We obtain an exact solution for $d \to \infty$
and a Hartree approximation for
finite $d$. They yield a Flory-like roughness exponent
$\zeta$ and a non trivial anomalous diffusion
exponent $\nu$ continuously dependent
on the ratio $g_{\rm T}/g_{\rm L}$ of divergence-free ($g_{\rm T}$) 
to potential ($g_{\rm L}$) disorder strength.
For the particle ($D=0$) our results
agree with previous order $\epsilon^2$ RG calculations.
The time-translational invariant
dynamics for $g_{\rm T} >0$ smoothly crosses over
to the previously studied ultrametric 
aging solution in the potential case.
\end{abstract}

The problem of the diffusion of a particle
in a $d$-dimensional static random force field has been often addressed
in the past decade
\cite{global,ledou_lrcorr,eps2,durrett,machta,ledou_1d}.
It is of interest
in problems like tracer diffusion in porous media \cite{global,havlin} or
turbulent flows \cite{hentschel}. Typically the
force is the sum of a potential random force (longitudinal in $k$-space)
of strength $g_{\rm L}$ and a divergence-free 
(transverse) one of
strength $g_{\rm T}$. The diffusion is anomalous
\cite{global,ledou_lrcorr,eps2,durrett,machta,ledou_1d}
($r \sim t^{\nu}$ with
$\nu \neq 1/2$) if either
(i) the dimension $d$ of space is low ($d \leq 2$), or
(ii) the force field has long-range correlations
($\overline{F(0) F(r)} \sim r^{-a}$ with $a \leq 2$).
While this is true for weak gaussian disorder, other anomalous 
behaviors are possible for disorder distributions with tails
\cite{machta}. Apart from particular 
cases (such as the long-range correlated potential case \cite{durrett},
including $d=1$~\cite{ledou_1d}),
most of the above results were obtained by perturbation theory
in the disorder around the thermal value $\nu_{\rm th}=1/2$,
which holds in the quadrant $d>2$ and $a>2$.
An expansion in powers of $\epsilon'= 2-d$ was performed for short range (SR)
disorder \cite{global}, and one in powers of $\epsilon= 2-a$ for long range (LR)
correlated disorder \cite{ledou_lrcorr,eps2}
(and a double $\epsilon$, $\epsilon'$ expansion \cite{eps2} near
the point $d=2$, $a=2$). Superdiffusive
behavior was obtained for $g_{\rm T}/g_{\rm L} \gg 1$ and subdiffusive
slow dynamics for $g_{\rm T}/g_{\rm L} \ll 1$. 
A {\it line\/} of fixed points was obtained
up to order $\epsilon^2$, with a continuously variable exponent $\nu$, in the
case of LR correlations ($d>2,$ $a<2$), parametrized
by the ratio $g_{\rm T}/g_{\rm L}$. The divergence-free case $g_{\rm L}=0$ is
simpler, since the value of $\nu$ is equal to the Flory
value $\nu_{\rm F}$ obtained by purely dimensional arguments, 
as discussed below.

On the other hand, out of equilibrium relaxational (potential)
dynamics has recently been studied extensively,
in the context of spin glass models \cite{Cuku}
as well as particles\cite{Frme,leto1} and manifolds \cite{leto2} in random potentials.
Analytical solutions were obtained \cite{Cuku} for the long-time dynamics
in mean-field models, showing explicitly their
glassy features such as slow (aging) dynamics, and
the breakdown of time-translation invariance (TTI)
and of fluctuation-dissipation theorems (FDT).
Whether these glassy properties 
of potential dynamics survive to a small {\it non-potential}
perturbation was recently addressed
\cite{leto4}, and it was found that, while
the dynamics becomes TTI in most cases, 
some glassy features survive.
Similarly, a glassy state with stationary dissipation
was proposed \cite{giamarchi_moving_prl}
for driven vortex systems.

In this paper we thus reconsider the dynamics of
a particle ($D=0$) in a $d$-dimensional random force field
as well as its extension to a directed elastic manifold of internal dimension
$D$.  We obtain the
diffusion exponent $r \sim t^{\nu}$ and (for $D>0$)
the roughness exponent $r \sim x^{\zeta}$ 
where $x$ is the internal coordinate.
Our results are exact for $d \to \infty$. This mean-field limit
is non trivial if transverse and longitudinal parts
are scaled properly. It extends the previous results 
\cite{ledou_lrcorr,eps2} obtained 
via RG since it allows to reach {\it nonperturbatively\/} 
arbitrary values of the ratio $g_{\rm T}/g_{\rm L}$ 
(that the RG could not reach
because of runaway flows for $g_{\rm T}=0$). We can now 
explore systematically the interesting potential limit
$g_{\rm T} \to 0$ and we prove 
the remarkable property of continuity of the 
mean-field dynamic solution
when aging is approached for $g_{\rm T} \to 0$.
This property was
proposed in \cite{leto4} and verified in
numerical solutions of the $p$-spin model.
We suggest to take systematically advantage of it, in order
to improve numerical simulations by adding a small nonpotential part,
waiting till TTI is attained, and taking the nonpotential part to
zero {\it after\/} the large size limit.
We find that the finite-$d$ Hartree
approximation compares well with the RG.

We study the Langevin dynamics of
a directed elastic manifold,
of internal dimension $D$, parametrized by a $d$-component field
$r(x)$:
\begin{equation} \label{depart}
\partial_t  r_{\alpha}(x,t)  = \nabla^2_x r_{\alpha}(x,t) + 
F_{\alpha}[r(x,t),x] + \eta_{\alpha}(x,t)\;,
\end{equation}
where $\eta$ is the gaussian thermal noise, satisfying
$\langle \eta_{\alpha}(x,t) \eta_{\beta}(x',t')  \rangle=
2 T \delta_{\alpha \beta} \delta(t-t') \delta^D(x-x')$, and
$F_{\alpha}[r,x]$ is the gaussian quenched random force
with a correlator 
$
\overline{F_{\alpha}[r,x]  F_{\beta}[r',x']} 
= \Delta_{\alpha \beta}(r-r') \delta^D(x-x')
$
which depends only on the distance $|r-r'|$.
We denote by angular brackets
averages over the thermal noise and by an overbar
averages over the random force. The force has both a potential (L)
and a divergence-free (T) part, which contribute separately to the correlator:
\begin{equation}
\Delta_{\alpha \beta}(r)  = -
\partial_{r_\alpha}  \partial_{r_\beta}  
\Delta_{\rm L} (r) -  \left( 
\delta_{\alpha \beta} \sum_{\gamma} \partial^2_{r_\gamma} 
 -  \partial_{r_\alpha}  \partial_{r_\beta} \right)
\Delta_{\rm T} (r)\;.
\end{equation}

We study the case in which the correlations at large scale behave
as power laws: the Fourier transforms of the correlators
with respect to $r$ behave, for small values of their argument $K$, as
$
\Delta_{\alpha \beta}(K) \sim K^{a-d} ( 
\tilde{g}_{\rm L} P^{\rm L}_{\alpha \beta}(K)
+ \tilde{g}_{\rm T} P^{\rm T}_{\alpha \beta}(K) )
$,
where $P_{\alpha\beta}^{\rm L}(K)=K_{\alpha}K_{\beta}/K^2$
is the longitudinal projector and $P_{\alpha\beta}^{\rm T}(K)=1-P_{\alpha
\beta}^{\rm L}(K)$ is the transverse one. 
We compute the mean-squared displacement $D_{xx'}(t,t')$
and the correlation $B_{xx'}(t,t')$, respectively defined by
$
D_{xx'}(t,t') = \frac{1}{d} 
\overline{\langle (r(x,t)-r(x',t'))^2 \rangle}
$ 
and
$
B_{xx'}(t,t') = \frac{1}{d} \overline{\langle
(r(x,t)-r(x,t'))\cdot (r(x',t)-r(x',t')) \rangle}
$.
The mean-square {\em displacement\/} at equal {\it times\/}
$D_{xx'}(t,t) \sim |x-x'|^{2 \zeta}$ measures the
roughness of the manifold,
while the {\em correlation\/} at equal {\it space\/}
$\tilde{B}(t,t') = B_{xx}(t,t')$ measures the
diffusion of a single monomer. When 
TTI holds,
the diffusion exponent $\nu$ is defined by
$\tilde{B}(t,t') \sim |t-t'|^{2 \nu}$.
In the absence of disorder one has
$\zeta=\zeta_{\rm th}=(2-D)/2$, $\nu=\nu_{\rm th} = (2-D)/4$
for $D<2$, and bounded motion and fluctuations
for $D>2$. 
Before proceeding, let us consider the problem by Flory-like
dimensional arguments.
If all terms in (\ref{depart}) scaled the
same way one would have $r/t \sim r/x^2 \sim r^{-\gamma} x^{-D/2}$ 
with $\gamma=a/2$ for $d > a$.
This yields
\begin{equation} \label{flory}
\zeta_{\rm F} = 2 \nu_{\rm F} = \frac{4-D}{4(1+\gamma)}\;.
\end{equation}
It turns out that, for $d \to \infty$,
$\zeta = \zeta_{\rm F}$ is exact
but $\nu = \nu_{\rm F}$ holds only in the transverse
case.

One can take the large-$d$ limit directly in the
De~Dominicis-Janssen 
functional, performing a gaussian decoupling
of $r(x,t)$ and the conjugate field
$\hat{r}(x,t)$. We define:
\begin{equation}
\Delta_{\rm T}(r) = -  V_{\rm T}(r^2/d)\;,\qquad
\Delta_{\rm L}(r) = -  d  \;  V_{\rm L} (r^2/d )\;.
\end{equation}

One can also study in any $d$ the self-consistent
Hartree approximation 
where $r(x,t)$ and $\hat{r}(x,t)$ are treated as
Gaussian variables. It amounts to replace 
everywhere, in the equations and solutions obtained below for $d=\infty$,
$V_{\rm L,T}$ by $\tilde{V}_{\rm L,T}$ defined by
$\tilde{V}(z) = \int d^{d} \phi/(2 \pi z)^{d/2} 
\exp(- \phi^2/(2z)) V(\phi^2/d)$
(see, e.g., appendix A of \cite{leto1}).
The coupled dynamical equations also involve the
response 
$
R^{\alpha \beta} _{xx'}(t,t') =  1/d \,  \overline{ 
\delta \langle r^{\alpha}(x,t) \rangle/\delta 
f^{\beta}(x',t') }|_{f=0}
$
where $f(x',t')$ is a small 
perturbation applied at point $x'$
at time $t'$. The response and
correlations are isotropic,
$R^{\alpha \beta}_{xx'}(t,t')
= (\delta_{\alpha \beta}/d) R_{xx'}(t,t')$. We
choose a space-homogeneous initial condition
$B_{x x'}(0,0) = B_{x-x'}(0,0)$ and denote by 
$B_k$ and $R_k$ the Fourier transforms
with respect to $x-x'$, and by tilde equal space
($x=x'$) two-time functions (e.g.,
$\tilde{B}=\int_k B_k$).
We use the shorthand $\int_k \equiv \int d^D k/(2\pi)^D $
and $\int_\omega \equiv \int d \omega/(2\pi)$.
With $R_k(t,t)=0$ and $R_k(t,t^-)=1$,
the dynamical equations read
\begin{eqnarray}
\partial_t R_k(t,t')
&=& 
- k^2 R_k(t, t') 
-  \int_0^t ds \; \Sigma(t,s)
\left( R_k(t,t') - R_k(s,t') \right)  + \delta(t-t')
\;,
\\
\label{qfulleqr} 
\partial_t B_k(t,t')
&=&
- k^2  \ B_k(t,t')
+
2 T
+ \, \int_0^t ds \;  D(t,s) \;  R_k(t,s) 
- \, \int_0^{t'} ds \;  D(t,s) \;  R_k(t',s) 
\nonumber\\
& &
- \,
\int_0^t ds \;  \Sigma(t,s) 
\; \left( B_k(t,s) - B_k(s,t') + B_k(t,t') \right)  - 4 T R_k(t',t)
\;  ,
\label{qfulleqb}
\end{eqnarray}
where $D(t,s) = 4 V_1'(\tilde B(t,s))$ and
$\Sigma(t,s) = - 4 V_2''(\tilde B(t,s)) \tilde R(t,s) $.
We define $V_1 = V_{\rm L} + (1 - 1/d)V_{\rm T} $ and $V_2 = V_{\rm L} $,
and consider
$ V_1(b) = g_1 b^{1-\gamma}/(2(1-\gamma))$, 
$V_2(b) = g_2 b^{1-\gamma}/(2(1-\gamma))$
where $g_1 = g_{\rm L} + (1-1/d) g_{\rm T}$, $g_2 = g_{\rm L}$,
and $\gamma = a/2$ for large $d$, thus generalizing the potential
case studied in \cite{leto1,leto2}.  We define
$\tilde{g}_{\rm L,T} = - g_{\rm L,T} C_{d,\gamma}/(2(1-\gamma))$, where
$C_{d,\gamma} = \int_r (r^2/d)^{1-\gamma} \exp(i n\cdot r)$
with $n^2=1$.  As shown below, the case of LR correlations
corresponds (for $d \to \infty$) 
to $\gamma > 2/(2-D)$ for $D<2$, and any $\gamma$ for $D>2$.

When $g_{\rm T} \neq 0$ one can assume that a TTI
solution is reached at large $t'$,
$B_k(t,t') = B_k(t-t')$ and $R_k(t,t') = R_k(t-t')$.
We checked this property numerically for $D=0$. However
we stress that it is not certain that it 
should hold for any nonpotential dynamics
\cite{footnote2}. Using TTI one can recast the above equations in the form
\begin{eqnarray}
R_k(\omega) = \frac{1}{ - i \omega + k^2 + \Sigma(0) - \Sigma(\omega)}
\;,
\;\;
 B_k(\omega) =  \frac{2 T + D_k}{k^2}  2 \pi \delta(\omega) 
- \frac{ 4 T + D(\omega) }{ | - i \omega + k^2 + \Sigma(0) - \Sigma(\omega) |^2}
\; ,
\nonumber
\\
\Sigma(\omega) = - 4 \int_{-\infty}^{+\infty} dt \, \exp(i \omega t) \, 
V_2''(b(t)) \, r(t) \;, 
\;\;\;\;\;\;\;\;
D(\omega) = 4 \int_{-\infty}^{+\infty} dt \, \exp(i \omega t) \,
V_1' (b(t)) 
 \label{tti}
\; ,
\end{eqnarray}
where $D_k = \int_ {\omega'} ( D(\omega') R_k(-\omega') -
\Sigma(\omega') B_k(\omega') )$. 
The correlation $C_k(\omega) =
1/d \overline{\langle (r(k,\omega).r(-k,-\omega)\rangle}$
thus obeys $2 C_k(\omega) = 
(4 T + D(\omega) )/| - i \omega
+ k^2 + \Sigma(0) - \Sigma(\omega) |^2$. 
The displacement can
then be computed using
$D_x(t) = 2 \int_{k,\omega} (1 - \cos(kx + \omega t))C_k(\omega)$.
In the large time limit we propose
\begin{equation} \label{asympt}
\tilde{R}(t) \sim  R \, t^{-\mu}\;,\qquad\qquad
\tilde{B} (t) \sim B \, t^{2 \nu}
\; ,
\end{equation}
and define the FDT violation factor $X[B]$
by $\tilde R(t) = X[\tilde{B}(t)] \partial_t \tilde{B}(t)$. 
Equation (\ref{asympt}) implies the small frequency behaviors
$\tilde R(\omega) \sim  R \Gamma[1-\mu] (- i\omega)^{\mu-1}$ for $\mu < 1$
or $\tilde R(\omega)-\tilde R(0) 
\sim  R \Gamma[1-\mu] (- i\omega)^{\mu-1}$ for $\mu  > 1$
and $\tilde B(\omega) 
\sim 2 B \Gamma[1+2 \nu]  \; {\mbox{Re}} (i\omega)^{-1-2 \nu}$.
It also implies $D(\omega) \sim  8 g_1 B^{-\gamma} \Gamma[1 - 2 \nu \gamma] \;
{\mbox {Re}} (i\omega)^{2 \nu \gamma -1}$ and
$\Sigma(0) - \Sigma(\omega)  \sim
- 4  g_2 \gamma R B^{-(1+\gamma)} \Gamma[1 - \mu - 2 \nu (\gamma+1)] 
 (- i\omega)^{\mu - 1 + 2 \nu (\gamma +1)} $.
One must have $0< \mu - 1 + 2 \nu (\gamma +1) <1$ in order to be able to neglect
the $- i \omega$ in the Green functions 
and to have a finite $\Sigma(0)$. We shall 
find out that this is always the case for $g_2 >0$. 
The equation for $\tilde{R}= \int_k R_k$ yields, at small $\omega$ and
for $D<2$,
\begin{equation}
\tilde R(\omega) \sim A_D ( \Sigma(0) - \Sigma(\omega))^{\frac{D-2}{2}} 
\; ,
\end{equation}
with $A_D = \pi/(2^D \pi^{D/2} \Gamma[D/2] \sin(\pi D/2))$. 
For $D>2$, $\tilde R(0)$ is finite and one has
instead $\tilde R(\omega) - \tilde R(0) \sim A_D 
( \Sigma(0) - \Sigma(\omega))^{\frac{D-2}{2}}$.
This gives the relation between exponents:
\begin{equation} \label{mu}
1 - \mu  = 2 \nu (\gamma + 1) \frac{2-D}{4-D}
\; .
\end{equation}
This value for $\mu$ is consistent with neglecting the $-i \omega$
term in (\ref{tti}), provided that
$0 < \nu < \nu_{\rm T}$, where $\nu_{\rm T} = (4-D)/(4 (1+ \gamma))$.
It implies $X[\tilde B] \sim X \tilde B^{w}$ with 
with $X=
R/(2 \nu) B^{-(1+\gamma) \frac{2-D}{4-D} }$
and $w= -1 + (1+\gamma) \frac{2-D}{4-D} $.
It also yields the following relation between amplitudes:
\begin{equation} \label{x}
(2 \nu X)^\frac{4-D}{2}  = 
A_D  \frac{ ( - 4  g_2 \gamma 
\Gamma[2 \nu (\gamma + 1) \frac{-2}{4-D} ] )^{\frac{D-2}{2}} }{
\Gamma[2 \nu (\gamma + 1) \frac{2-D}{4-D} ] }
\; .
 \end{equation}

Similarly the equation for $\tilde{B}= \int_k B_k$
yields
\begin{equation}
\tilde{B}(\omega) \sim A_D D(\omega) 
| \Sigma(0) - \Sigma(\omega) |^{\frac{D-4}{2}}
\frac{\sin(\phi \frac{2-D}{2})}{
\sin(\phi)}
\; ,
\end{equation}
where $\phi= - \frac{\pi}{2} ( \mu - 1 + 2 \nu (\gamma +1))$ is the
phase of $\Sigma(0) - \Sigma(\omega) $.
One can check that the exponent relation 
(\ref{mu}) is now automatically satisfied.
The resulting relation between amplitudes, together with the relation 
(\ref{x}), yields our main result, i.e., 
the equation which determines the exponent $\nu$:
\begin{equation} \label{nu}
\frac{g_2}{g_1} = h(\nu,D,\gamma) \equiv
\frac{\Gamma[2 \nu (\gamma + 1) \frac{2-D}{4-D} ] }
{|\Gamma[2 \nu (\gamma + 1) \frac{-2}{4-D} ] |}
\frac{  \Gamma[1 - 2 \gamma \nu] \sin(\pi \gamma \nu) 
}{ \Gamma[1 + 2 \nu]
\gamma \sin(\pi \nu) }
\frac{ \sin(\frac{\pi}{2} 2 \nu (1+\gamma) \frac{2-D}{4-D} ) }
{ \sin(\frac{\pi}{2} 2 \nu (1+\gamma) \frac{2}{4-D} ) }
\; ,
\label{nuRG}
\end{equation}
where, we recall, $g_2/g_1 = g_{\rm L}/(g_{\rm L}+g_{\rm T})$ when 
$d \to \infty$. The function
$h$ is a monotonic function of $\nu$, such that 
$h(\nu=0,D,\gamma)=1$ and $h(\nu = \nu_{\rm F},D,\gamma)=0$ where
$\nu_{\rm F}$ is given by (\ref{flory}). Thus we find that in the
transverse case $\nu = \nu_{\rm T} \equiv \nu_{\rm F}$ and that $\nu$
spans the interval $0< \nu < \nu_{\rm T}$ when the ratio $g_2/g_1$ varies from 
$1$ (potential case) to $0$ (divergence free case). These
results hold in the LR case $\gamma<\gamma_{\rm c}$.
The limit $\gamma \to \gamma_{\rm c}^{-}$ (for which $h(\nu,D,
\gamma_{\rm c})=1$)
is studied below. One also easily obtains the roughness exponent $\zeta$.
From $C_k(\omega)$ 
and Eq.~(\ref{tti}) one finds that $t \sim x^z$ with
$z=2 \nu/\nu_{\rm T}$ and $r \sim t^{\nu}$. Thus $\zeta=z \nu
=2 \nu_{\rm T} = \zeta_{\rm F}$.
This is the Flory value, as announced, and interestingly, is found not
to depend on the precise nature of the random force
(though the physics is quite different as the ratio $g_{\rm L}/g_{\rm T}$
varies). For finite $d$ our approach also yields a Hartree
approximation which amounts to: (i) replace in (\ref{nu})
$g_1$ by $ g_{\rm L} + (1-\frac{1}{d}) g_{\rm T}$ and $g_2$
by $g_{\rm L}$;
(ii) setting $\gamma=\min(a,d)/2$ ($\gamma = d/2$
corresponds to the Hartree SR case and
$\gamma < d/2$ the Hartree LR case since the model studied here and in 
\cite{ledou_lrcorr,eps2} is a random {\it force} model.

Let us compare our result for $\nu$ (exact when $d\to\infty$)
with the RG calculations for the particle ($D=0$) with 
$\gamma=1 - \epsilon$ in any $d$. The RG calculation was carried
out upto order $\epsilon$ in \cite{ledou_lrcorr}
and upto order $\epsilon^2$ in \cite{eps2}. In our notation, 
it reads \cite{eps2}
\begin{equation}
\frac{1}{2 \nu} = 1 -
\frac{d - 1 - d k}{2 (d - 1)} \epsilon
+ \epsilon^2 d k
\frac{
4 + 2 d - 5 d^2  + d^3  
+ d k( 4 - 2 d + d^2) }
{ 4 (2 - d) (d-1)^2  d } 
\; ,
\label{nud}
\end{equation}
where $k=g_L/g_T$, while, from (\ref{nu}) and using the above Hartree 
values one finds
\begin{equation}
\frac{1}{2 \nu} = 1 - \frac{g_1 - 2 g_2}{2(g_1-g_2)} \epsilon
- \frac{g_2 g_1 \epsilon^2}{4 (g_1-g_2)^2}
= 1 -
\frac{d - 1 - d k}{2 (d - 1)} \epsilon
+ \frac{\epsilon^2 d k (1 - d - d k)}{4 (d-1)^2}
\; .
\end{equation}
For $d \to \infty$, $\nu$  from (\ref{nuRG}) and (\ref{nud}) coincide.
$\nu$ is determined by the bare ratio $g_T/g_L$, which
indeed is unrenormalized in the RG.
Furthermore the finite $d$ Hartree approximation is
found to be {\it exact to first order} in $\epsilon=1-\gamma$ for all
$d>2$. Thus it is a good approximation in the Hartree LR case
and is exact for $d=1$ since it yields $\nu=0$.
Though it does predict correctly the special point 
$d=2, \gamma=1$ it is not able to reproduce correctly the delicate
crossover \cite{eps2} around this point. The RG approach of 
\cite{eps2,ledou_lrcorr} can be generalized to the manifold,
and for $0<D<2$ one can similarly expand around
$\gamma_c(D) = 2/(2-D)$ (LR) (and around 
$d_c(D)= 4/(2-D)$ (SR)). The result (\ref{nu})
leads to the following formula, 
exact for $d \to \infty$, with 
$\gamma=2/(2-D) - \epsilon$ to first order in $\epsilon$:
\begin{equation}
\nu = \frac{2-D}{4} - \epsilon
\frac{(2-D)^2}{8 (4-D)}
\frac{(4-D) g_2 - 2 g_1}{ g_1 - g_2 }\;.
\end{equation}
It would be interesting to check whether the
Hartree approximation is also exact for any $D<2$, $d$ to $O(\epsilon)$.

Up to now we have studied the limit $\gamma \to \gamma_c(D)$ 
keeping the ratio $g_{\rm L}/g_{\rm T}>0$ fixed, 
in which case $\nu \to \nu_{\rm th}$.
In fact the limit $\gamma \to \gamma_{\rm c}$ is very non uniform,
and therefore much richer. For $D<2$ one finds that
any value of $\nu$ with $0<\nu<\nu_{\rm th}$ can be reached
if one takes the limit $\gamma \to \gamma_{\rm c}(D)$ with the ratio
$g_{\rm T}/\epsilon g_{\rm L}$ fixed. For the particle, this is perfectly
consistent with the RG
result obtained \cite{eps2,ledou_lrcorr}
in any $d$, which in our notations read
$1/\nu = 1 + g_{\rm L} + O(g_{\rm L}^3)$
(note that it has a nice $d \to \infty$ limit and that 
there is a conjecture \cite{eps2,ledou_lrcorr} 
that all higher order terms vanish).
For $2<D<4$ one finds that the
limit $\gamma \to \infty$ is non trivial with $\nu \gamma$
a function of $g_{\rm T}/g_{\rm L}$. Finally the limit $D=4 - \epsilon$
is also very non uniform
and any value of $\nu$ can be reached if $g_{\rm T}/\epsilon$
is kept fixed as $\epsilon \to 0$. 

We now study the potential limit $g_{\rm T} \to 0$ and show on the case $D=0$
that one goes continuously to the aging dynamics which holds for
$g_{\rm T} = 0$. This is best illustrated using the formalism 
of Ref.~\cite{leto2}.
Both in the case of a solution satisfying ($g_{\rm T} > 0$) and in 
the asymptotic aging solution 
($g_{\rm T}=0$), one defines the FDT 
violation factor $X[B]$ and the triangular relation
$\overline{f}(B,B')$ by
\begin{eqnarray}
X[B] = \lim_{{t \geq t' \to \infty}\atop{ B(t,t')=B} } 
R(t,t')/\partial_{t'} B(t,t')
\; , 
\;\;\;\;\;\;\;\;\;\;\;
\overline{f}(B',B) = \lim_{{t \geq  s \geq t' \to \infty}\atop{B(t,t')=B 
\geq B(t,s)=B' }} B(s,t')
\; .
\end{eqnarray}
The dynamical equations are naturally written in terms of 
the linear susceptibility $\chi(t,t') = \int_{t'}^{t}  ds R(t,s)$,
 which takes the
form  $\chi( B) = \int_{0}^{ B} d B' \; X( B')$,  and of the ``anomaly''
$M[ B]  = 4 \int_{0}^{ B} d B' \; V''( B') X( B')$.  The equation for $ R$
only contains $V_2$ and is thus identical to the case $g_{\rm T}=0$ and
reads, in the long time limit, like Eq.~(5.3)  of Ref.~\cite{leto1}:
\begin{eqnarray}
0 \approx \frac{d \chi[ B] }{dt} &=& -1 
- 
\chi[ B]  \int_{ B}^{\infty} d B' \, 4 V_2''( B') \, X[ B'] 
+\int_{0}^{ B} d B' \, 4 V_2''( B') \, X[ B'] \, 
( \chi[\overline{f}( B', B)] - \chi[ B] )\;.
\end{eqnarray}
We used $B(t,0) \to \infty$.  
This equation is coupled to a similar
(more cumbersome) equation for $B$ that also involves $V_1$. 
We specialize to the long range case $\gamma < 1$. 
The integrals are naively divergent at large $ B$ since 
$\lim_{ B \to \infty} \chi[ B] = \infty$ \cite{footnote1}.
In the equation for $ B$ similar divergences occur.
However, at large $ B$, one finds
$\chi[ B] \sim 2 X/(1+\gamma)   B^{(1+\gamma)/2}$, as
well as
$\overline{f}( B', B) = ( B^{\frac{1}{2 \nu}} -  B'^{\frac{1}{2 \nu}})^{2 \nu}$.
The divergent parts cancel and one recovers the equation for the amplitudes
(\ref{x}) (independent of small $ B$ details).
This solution interestingly shows that
when $\nu \to 0$ as $g_T \to 0$ the triangular 
relation continuously becomes $\overline{f}(B',B) = \max(B',B)$, i.e.
the known ultrametric solution for the potential case. Similarly
$X[ B] \sim X  B^{(\gamma-1)/2}$ tends continuously to the result of \cite{leto1}.

In the transverse case $g_{\rm L}=0$ for the
particle $D=0$ our method yields the Flory estimate
$\nu=\nu_{\rm F} \equiv 1/(1 + \min(a,d)/2)$, which turns
out to be exact (for $\nu_{\rm F} > 1/2$).
This can either be seen perturbatively
to all orders in the field theory \cite{eps2},
or, more physically, using that the fact that 
the {\it stationary\/}
measure is exactly given by
$P_{\rm stat} ={\rm  const.} $ The
Derrida-Luck method \cite{derrida_luck,footnote4}
then produces the result (and also yields
rigorously the velocity-force law $v \sim F^{\phi}$ with $\phi=1$) as 
we now show.
One introduces a medium with random forces, but periodic
with a large period $\Lambda$.
An {\it exact\/} formula for the velocity in
this medium is $v_{\Lambda} = \int dx F(x) P_{\rm stat}$ where 
$P_{\rm stat}(x)$ is a stationary measure (satistying 
$0 = \nabla^2 P_{\rm stat} - \partial_i ( F_i P_{\rm stat}) )$ 
with periodic boundary conditions.
Using that $P_{\rm stat}=1/\Lambda^d$ one finds that the 
mean-squared velocity $\overline{ v_\Lambda^2 } \sim \Lambda^{-\min(a,d)}$
in that periodic medium (equivalently
finite torus of volume $\Lambda^d$).
This exact scaling relation ensures that $\nu = \nu_{\rm F}$.
For the manifold $D>0$, however, the stationary measure
is not known (the naive extension $P_{\rm stat} \sim
\exp( - \int dx (\nabla r)^2 /2 T)$ is not stationary).
Thus the Flory values obtained here for the manifold
in a divergence free flow
may very well be exact only for $d \to \infty$.

One can also extend the arguments of Ref.
\cite{hentschel} to the diffusion of an oriented (e.g.\ by an external field)
manifold in a divergence-free turbulent flow. If
the random force, proportional to
the local velocity, can be
considered as static on some time scales, 
with LR gaussian
correlations $\overline{v(0) v(r)} \sim r^{2/3 + \mu/3}$
($\mu$ arises from the intermittency corrections to
Kolmogorov scaling), the result $\nu = \nu_{\rm T}$ for the corresponding
diffusion exponent yields $\nu = 3 (4-D)/(2 (4-\mu))$,
which can be considered as a generalization of the
Richardson's law (corrected by intermittency).

In conclusion, we have studied the problem
of the dynamics of a manifold or a particle in a gaussian
random force flow. We have obtained the
roughness exponent $\zeta$ and $\nu$ exactly for
$d \to \infty$. Our study establishes a bridge between several
problems (aging and nonpotential dynamics) 
and techniques (large $d$ and RG approaches).

\vspace{.5cm}

We thank J. Kurchan for useful discussions.
LFC thanks the SPHT-Saclay for hospitality and the 
European Union  for financial    
support through the contract ERBCHRXCT920069.
LP is Associato INFN, Sezione di Napoli, and acknowledges the
support of a Chaire Joliot de l'ESPCI.


\begin{thebibliography}{99}
\bibitem{footnote1} This can be seen from Ref \cite{leto1} by letting 
$\mu \to 0$ since with a mass $\mu$ one has $\chi[B[t,0]] = 1/\mu$.
\end{thebibliography}
\end{document}